\documentclass[twocolumn,english]{article}
\usepackage[T1]{fontenc}
\usepackage[latin9]{inputenc}
\usepackage[a4paper]{geometry}
\usepackage{amstext}
\usepackage{amssymb}
\usepackage{graphicx}

\makeatletter

\providecommand{\tabularnewline}{\\}

\newcommand{\lyxaddress}[1]{
	\par {\raggedright #1
	\vspace{1.4em}
	\noindent\par}
}

\usepackage{xcolor}            
\definecolor{myred}{rgb}{0.66, 0.15, 0.15}

\usepackage{hyperref}          
\hypersetup{
     colorlinks   = true,
     citecolor    = myred,
     urlcolor     = myred,
     urlbordercolor = myred
}
\usepackage{cuted}

\makeatother

\usepackage{babel}
\begin{document}
\begin{strip}\vspace{-2cm}

\title{\textbf{Magnetic Feshbach resonances in $^{7}\text{Li}-$$^{133}\text{Cs}$
mixtures}}
\author{Pascal Naidon\\
{\footnotesize{}Strangeness Nuclear Physics Laboratory, RIKEN Nishina
Centre, Wak{\=o}, 351-0198 Japan. }}
\maketitle

\lyxaddress{\vspace{-0.5cm}
}
\begin{abstract}
Motivated by the prospect of observing Efimov and Bose polaron physics
in ultracold mixtures of bosonic atoms with large mass imbalance,
this work investigates the magnetic Feshbach resonances between $^{7}\text{Li}$
and $^{133}\text{Cs}$. The resonances are predicted at the 1 gauss
level using the model of Ref.~\cite{Pires2014a} obtained from experimental
observations of resonances between $^{6}\text{Li}$ and $^{133}\text{Cs}$.
It is found that a few resonances in a practical range of magnetic
field intensity could be used to tune the scattering length between
$^{7}\text{Li}$ and $^{133}\text{Cs}$ atoms. Opportunities for observing
Efimov and Bose polaron physics are discussed.

\vspace{0.5cm}
\end{abstract}
\end{strip}

\section{Introduction}

The Efimov effect has been studied experimentally for several years
with ultracold atoms. The effect is enhanced for systems of two heavy
and one light particles, near the two-body resonance between a heavy
and a light particle. This has been observed in mixtures of atoms
with a large mass imbalance such as $^{6}\text{Li}$ and $^{133}\text{Cs}$,
for which the universal scaling factor between subsequent three-body
states is 4.88 instead of 22.7 for three particles of identical mass,
enabling the observation of up to three Efimov three-body bound states~\cite{Tung2014,Pires2014}.
The experiments are performed at sufficiently low density to assume
in first approximation that the three-body observables are unaffected
by the surrounding medium. However, it was found theoretically that
the medium can make significant changes to the Efimov three-body spectrum
and can even lead to an interesting interplay between Efimov and polaron
physics. In particular, the case of heavy impurities immersed in a
Bose-Einstein condensate of light atoms leads to a crossover between
a phonon-mediated Yukawa force and particle-mediated Efimov attraction~\cite{Naidon2016b},
leading to the prospect of observing bipolarons and tripolarons in
the crossover region where the mediated interaction becomes resonant.

One of the candidate species to observe this physics are $^{133}\text{Cs}$
atoms immersed in a Bose-Einstein condensate of $^{7}\text{Li}$ atoms.
To this purpose, the use of a two-body resonance between these two
species is needed. While the magnetic Feshbach resonances between
$^{7}\text{Li}$ and $^{133}\text{Cs}$ have not been observed yet,
the resonances of $^{6}\text{Li}$ and $^{133}\text{Cs}$ have been
precisely measured and theoretically modelled~\cite{Repp2013,Tung2013,Pires2014a}.
The purpose of this work is to use this theoretical model to obtain
predictions of the Feshbach resonances of $^{7}\text{Li}$ and $^{133}\text{Cs}$,
as a guide for experimental investigation.

\begin{table*}
\begin{centering}
\begin{tabular}{|cccccc|}
\hline 
Entrance channel & Experiment~\cite{Tung2013} & Theory~\cite{Tung2013} & Experiment~\cite{Pires2014} & Theory~\cite{Pires2014a} & This work\tabularnewline
\hline 
$^{6}$Li$\vert1/2,+1/2\rangle\oplus$ $^{133}$Cs$\vert3,+3\rangle$ & 843.4(2) & 843.1(2) & 843.5(4) & 842.99 & 843.80\tabularnewline
 & 892.9(2) & 893.0(2) & 892.87(7) & 892.98 & 893.98\tabularnewline
\hline 
$^{6}$Li$\vert1/2,-1/2\rangle\oplus$ $^{133}$Cs$\vert3,+3\rangle$ & 816.1(2) & 816.4(2) & 816.24(2) & 816.36 & 817.05\tabularnewline
 & 889.0(2) & 888.8(2) & 889.2(2) & 888.74 & 889.56\tabularnewline
 & 943.4(2) & 943.4(2) & 943.26(3) & 943.38 & 944.39\tabularnewline
\hline 
\end{tabular}
\par\end{centering}
\caption{\label{tab:Positions6Li-133Cs}Positions (magnetic field in gauss)
of resonances between $^{6}\text{Li}$ and $^{133}\text{Cs}$, for
different hyperfine entrance channels labelled by the hyperfine state
$\vert f,m_{f}\rangle$ of each atom.}
\end{table*}
\begin{figure*}
\centering{}\includegraphics[width=9cm]{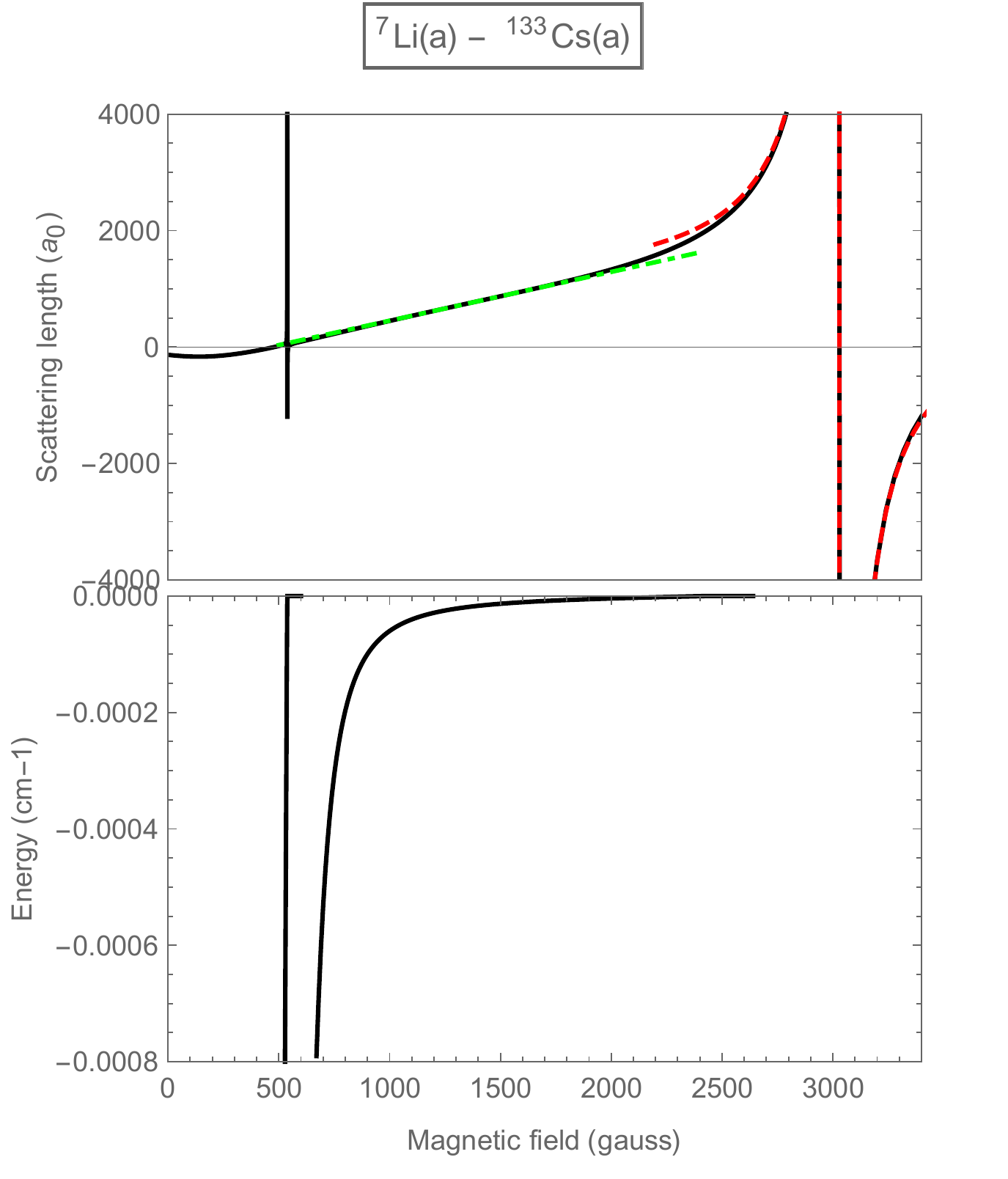}\includegraphics[width=9cm]{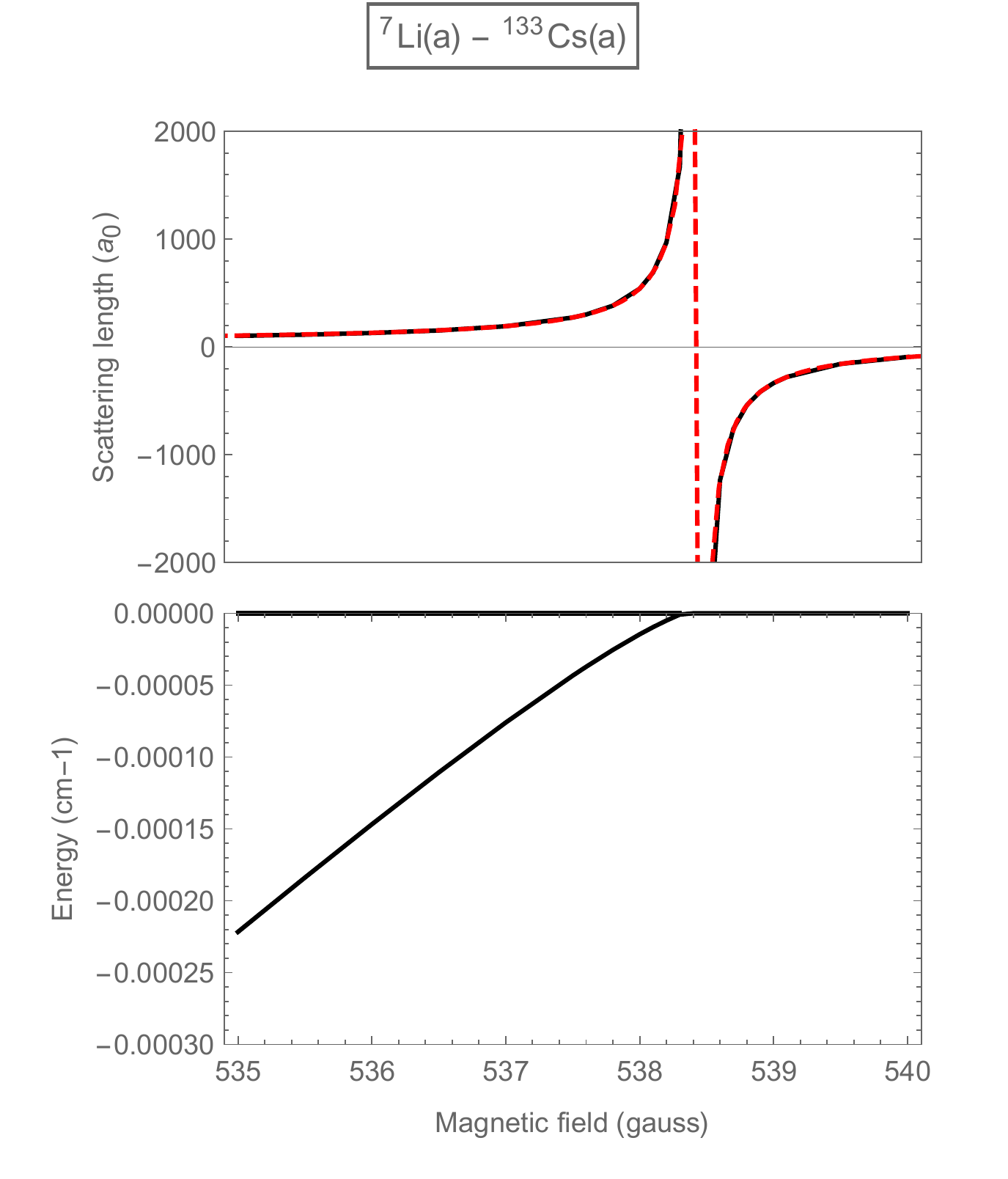}\caption{\label{fig:aa}Scattering length and near-threshold bound states in
the $aa$ entrance channel. The right panel is a close-up near the
first resonance. In each panel, the red dashed curve represents the
resonance formula Eq.~(\ref{eq:ResonanceFormula}) with its parameters
given in Table~\ref{tab:Positions7Li-133Cs}. The green dot-dashed
line in the left panel represents the linear fit: $a=0.8382(B-460.237)$.}
\end{figure*}
\begin{figure}
\centering{}\includegraphics[scale=0.6]{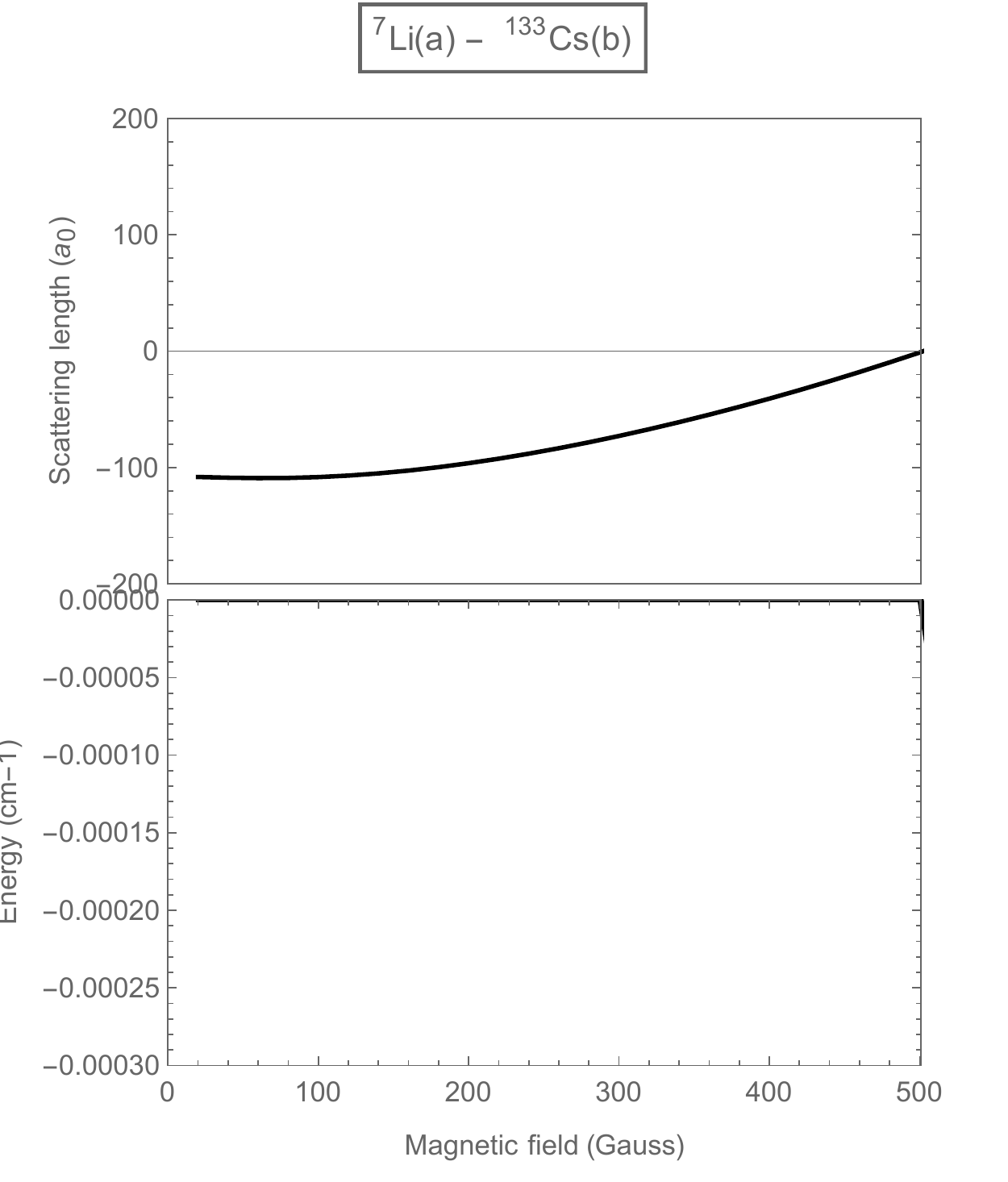}\caption{\label{fig:ab}Scattering length and near-threshold bound states (none
visible in the plotted energy range) in the $ab$ entrance channel.}
\end{figure}
\begin{figure*}
\centering{}\includegraphics[width=9cm]{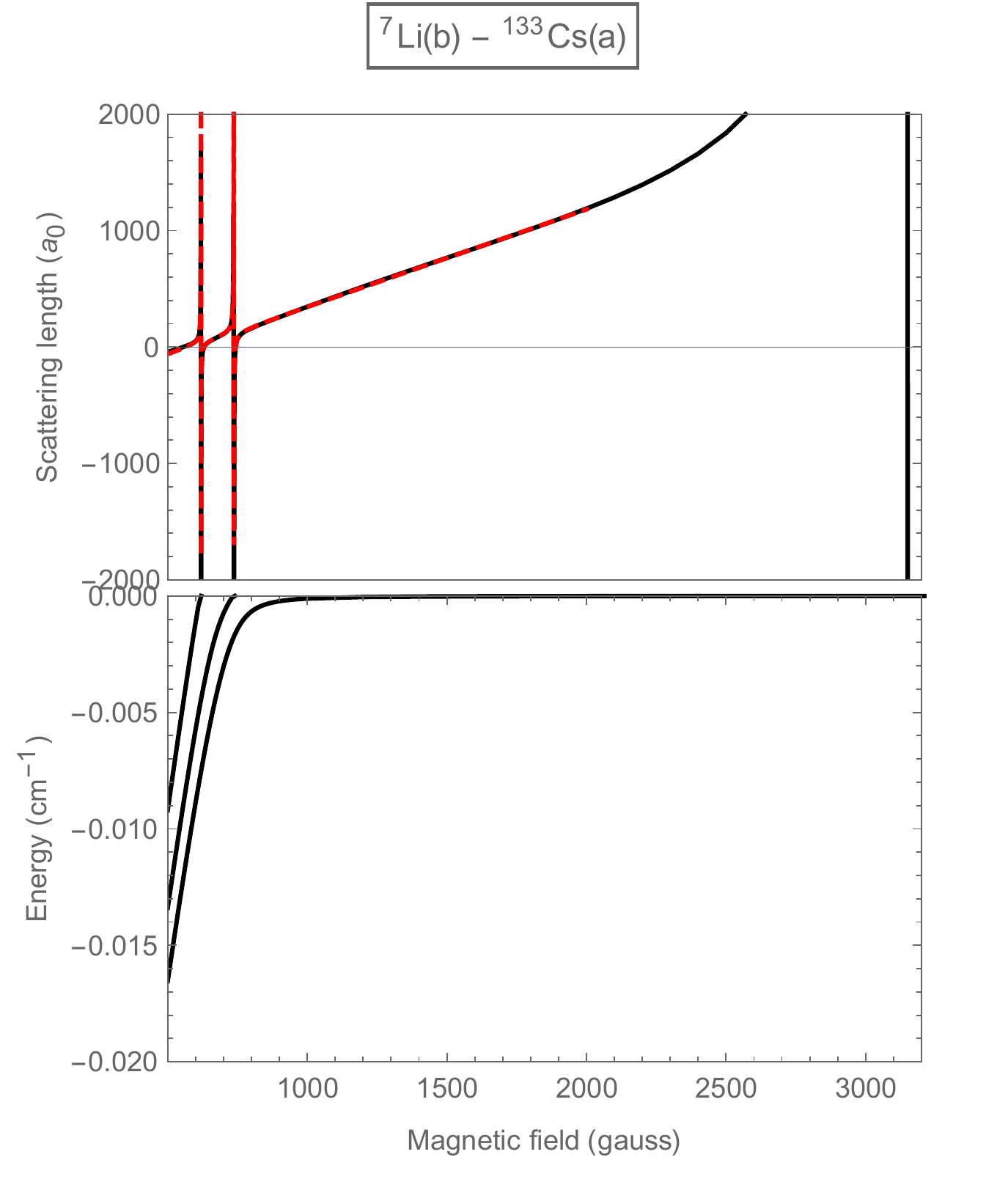}\includegraphics[width=9cm]{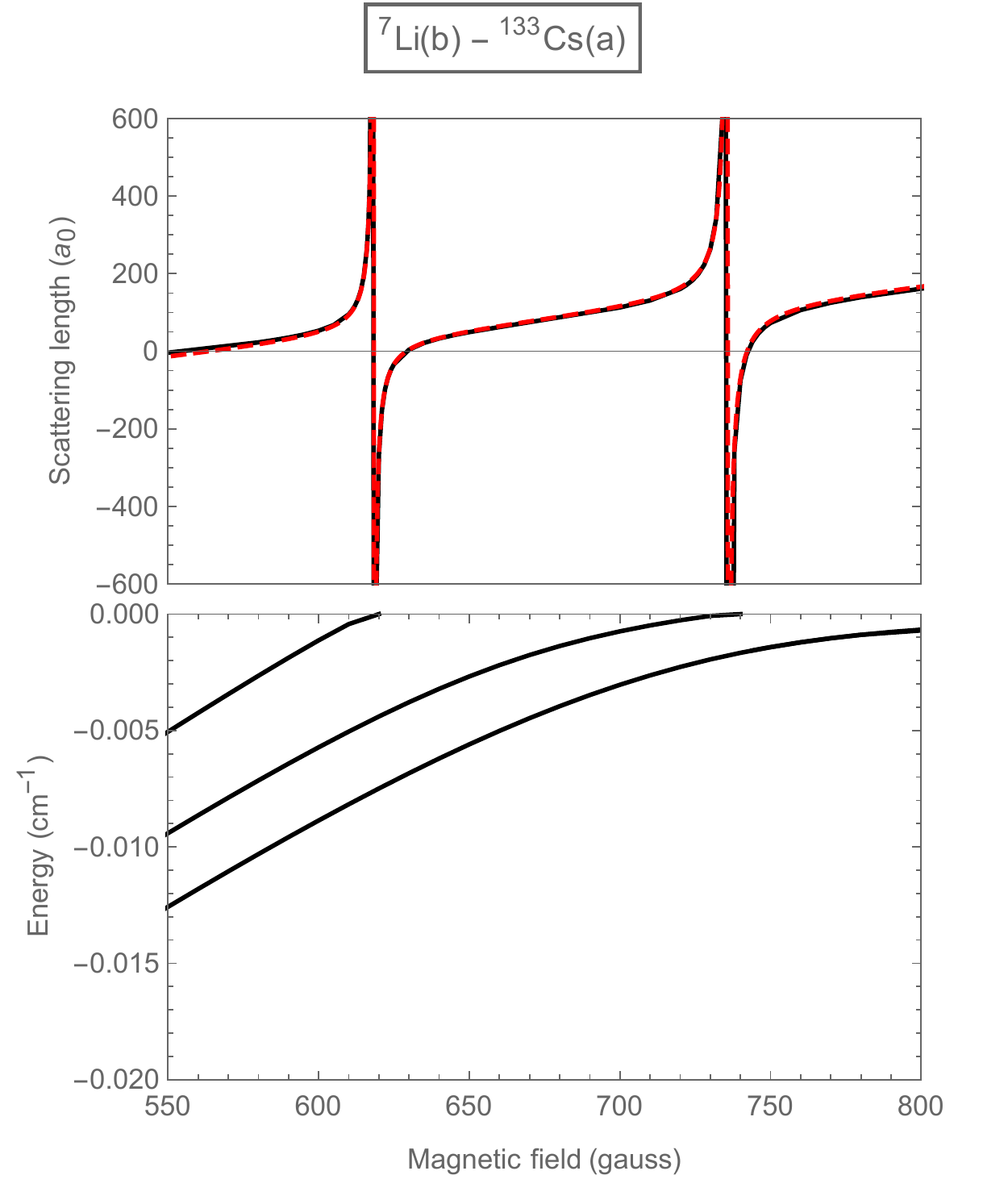}\caption{\label{fig:ba}Scattering length and near-threshold bound states in
the $ba$ entrance channel. The right panel is a close-up near the
first resonances. In both panels, the red dashed curve represents
the fit formula: $a=0.833636(B-580.992)-\Delta^{(1)}/(B-B_{0}^{(1)})-\Delta^{(2)}/(B-B_{0}^{(2)})$
where the resonance parameters $\Delta$ and $B_{0}$ are given in
Table~\ref{tab:Positions7Li-133Cs}.}
\end{figure*}

\begin{table*}
\begin{centering}
\begin{tabular}{|cccc|}
\hline 
Entrance channel & $B_{0}$~(G) & $\Delta$~(G) & $a_{\text{bg}}$~($a_{0}$)\tabularnewline
\hline 
$^{7}$Li$\vert1,+1\rangle\oplus$ $^{133}$Cs$\vert3,+3\rangle$
$\quad(aa)\quad$ & 538.43 & 4.89 & 43.85\tabularnewline
 & 3029.95 & 918.74 & 838.95\tabularnewline
\hline 
$^{7}$Li$\vert1,0\rangle\oplus$ $^{133}$Cs$\vert3,+3\rangle$ $\quad(ba)\quad$ & 618.31 & 13.29 & 39.25\tabularnewline
 & 735.76 & 6.37 & 132.36\tabularnewline
 & 3149.85 & 874.77 & 840.645\tabularnewline
\hline 
\end{tabular}
\par\end{centering}
\caption{\label{tab:Positions7Li-133Cs}Predicted resonances between $^{7}\text{Li}$
and $^{133}\text{Cs}$, for different hyperfine entrance channels
labelled by the hyperfine state $\vert f,m_{f}\rangle$ of each atom.}
\end{table*}

\section{Model}

The model used in this work is based on the standard two-atom Hamiltonian,
\begin{equation}
\hat{H}=\hat{T}+\hat{V}+\hat{H}_{A}+\hat{H}_{B},\label{eq:MolecularHamiltonian}
\end{equation}
where $\hat{T}=-\hbar^{2}\nabla_{R}^{2}/(2\mu)$ is the relative kinetic
energy of the two atoms, with a reduced mass $\mu$, and $\hat{V}$
is the interaction potential between the two atoms, assumed to depend
only on the total electronic spin and relative distance $R$ of the
two atoms. The Hamiltonians $\hat{H}_{A}$ and $\hat{H}_{B}$ of each
separated atom $A$ and $B$, consist of hyperfine and Zeeman terms
and read (for $k=A,B$),
\begin{equation}
H_{k}=\underbrace{A_{k}\hat{\vec{s}_{k}}\cdot\hat{\vec{i}}_{k}}_{\mbox{Hyperfine}}+\underbrace{(g_{j,k}\hat{\vec{s}}_{k}+g_{i,k}\cdot\hat{\vec{i}}_{k})\cdot\mu_{B}\vec{B}}_{\mbox{Zeeman}}\label{eq:AtomicHamiltonian}
\end{equation}
where $\hat{\vec{s}_{k}}$ and $\hat{\vec{i}}_{k}$ are respectively
the electronic and nuclear spins of atom $k$, $A_{k}$ is its hyperfine
structure constant, and $g_{j,k}$ and $g_{i,k}$ are its electronic
and nuclear gyromagnetic factors. Here, we have neglected the electronic
spin-spin interaction and the effective $R$-dependence of the hyperfine
structure constants $A_{k}$. These effects, which were included in
the model of Ref.~\cite{Pires2014a}, are expected to affect the
positions of resonances by about a gauss or less. Indeed, solving
the above model for $A=\;^{6}\text{Li}$ and $B=\;^{133}\text{Cs}$
with the singlet and triplet potentials given in Ref.~\cite{Pires2014a},
yields all the resonance positions of that system within one gauss
of the reported experimental and theoretical values, as shown in Table~\ref{tab:Positions6Li-133Cs}.

In the absence of experimental data for $A=\;^{7}\text{Li}$ and $B=\;^{133}\text{Cs}$,
one can assume that the same singlet and triplet potentials as those
for $^{6}\text{Li}$ and $^{133}\text{Cs}$ can be used. This approximation
justifies to limit the accuracy of our model to the one gauss level.
Due to the isotopic mass difference, the singlet and triplet scattering
lengths $a_{s}$ and $a_{t}$ change from $(a_{s},a_{t})=(30.15,-34.24)\,a_{0}$
for $^{6}\text{Li}$-$^{133}$Cs, to $(a_{s},a_{t})=(45.47,908.2)\,a_{0}$
for $^{7}$Li-$^{133}$Cs, as already noted in Ref.~\cite{Repp2013}\footnote{The values of $a_{s}$ and $a_{t}$ obtained from the potentials of
Ref.~\cite{Pires2014a} are slightly different from the values stated
in that reference and Ref.~\cite{Pires2014}, namely (30.252(100),-34.259(200))
and (45.477(150),908.6(100)), although within the theoretical uncertainty.
They are also close to the values (30.2(1),-34.5(1)) reported in Ref.~\cite{Tung2013}.}. This makes the triplet scattering length of this system atypically
large compared to the mean scattering length $\bar{a}=44.40\,a_{0}$.
(Here, $a_{0}$ denotes the Bohr radius).

The remaining difference concerns the hyperfine structure. Lithium-7
has a nuclear spin 3/2 and electronic spin 1/2, resulting in a total
spin $f_{A}=1,2$. We enumerate the corresponding 3+5 hyperfine states
from $a$ to $h$. Caesium-133 has a nuclear spin 7/2, resulting in
a total spin $f_{B}=3,4$. We enumerate the corresponding 7+9 hyperfine
states from $a$ to $p$. Since the total projection $m_{F}=m_{f_{A}}+m_{f_{B}}$
is conserved during a scattering of the two atoms, and since we are
interested in elastic collisions, we consider the ground-state entrance
channel of a given $m_{F}.$ For $m_{F}=4$, the ground-state entrance
channel corresponds to the two atoms in their ground states, $^{7}\text{Li}\vert1,+1\rangle$
and $^{133}\text{Cs}\vert3,+3\rangle$ (the $aa$ channel). For $m_{F}=3$,
the ground state entrance channel corresponds to $^{7}\text{Li}\vert1,+1\rangle$
and $^{133}\text{Cs}\vert3,+2\rangle$ (the $ab$ channel) for $B<B_{ab}$,
and $^{7}\text{Li}\vert1,0\rangle$ and $^{133}\text{Cs}\vert3,+3\rangle$
(the $ba$ channel) for $B>B_{ab}$, where $B_{ab}=501.024$G is the
magnetic field at which the $ab$ channel crosses the $ba$ channel.

\section{Results}

\begin{figure*}
\begin{centering}
\includegraphics[viewport=50bp 0bp 480bp 328bp,clip,width=9cm]{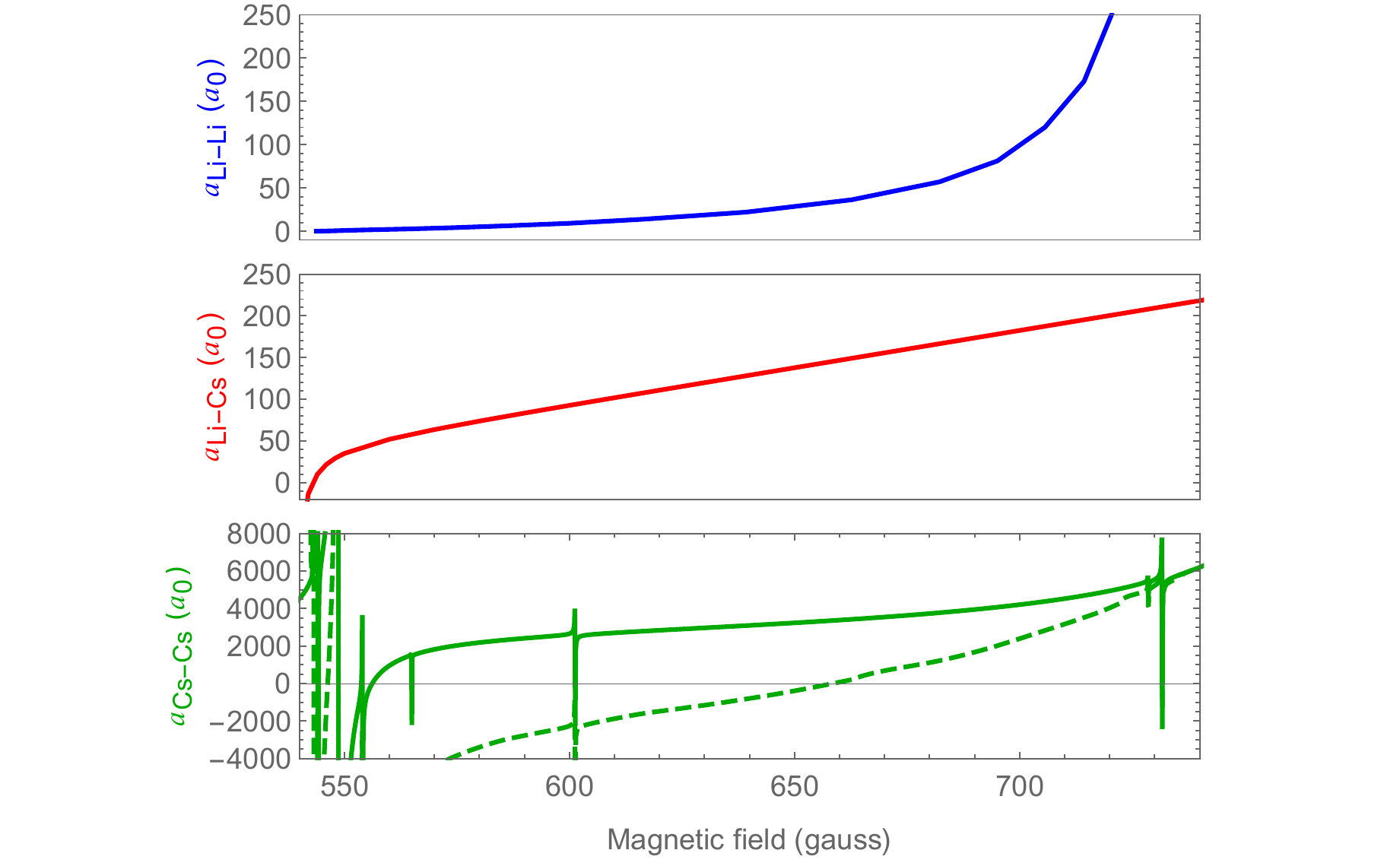}\includegraphics[viewport=50bp 0bp 480bp 328bp,clip,width=9cm]{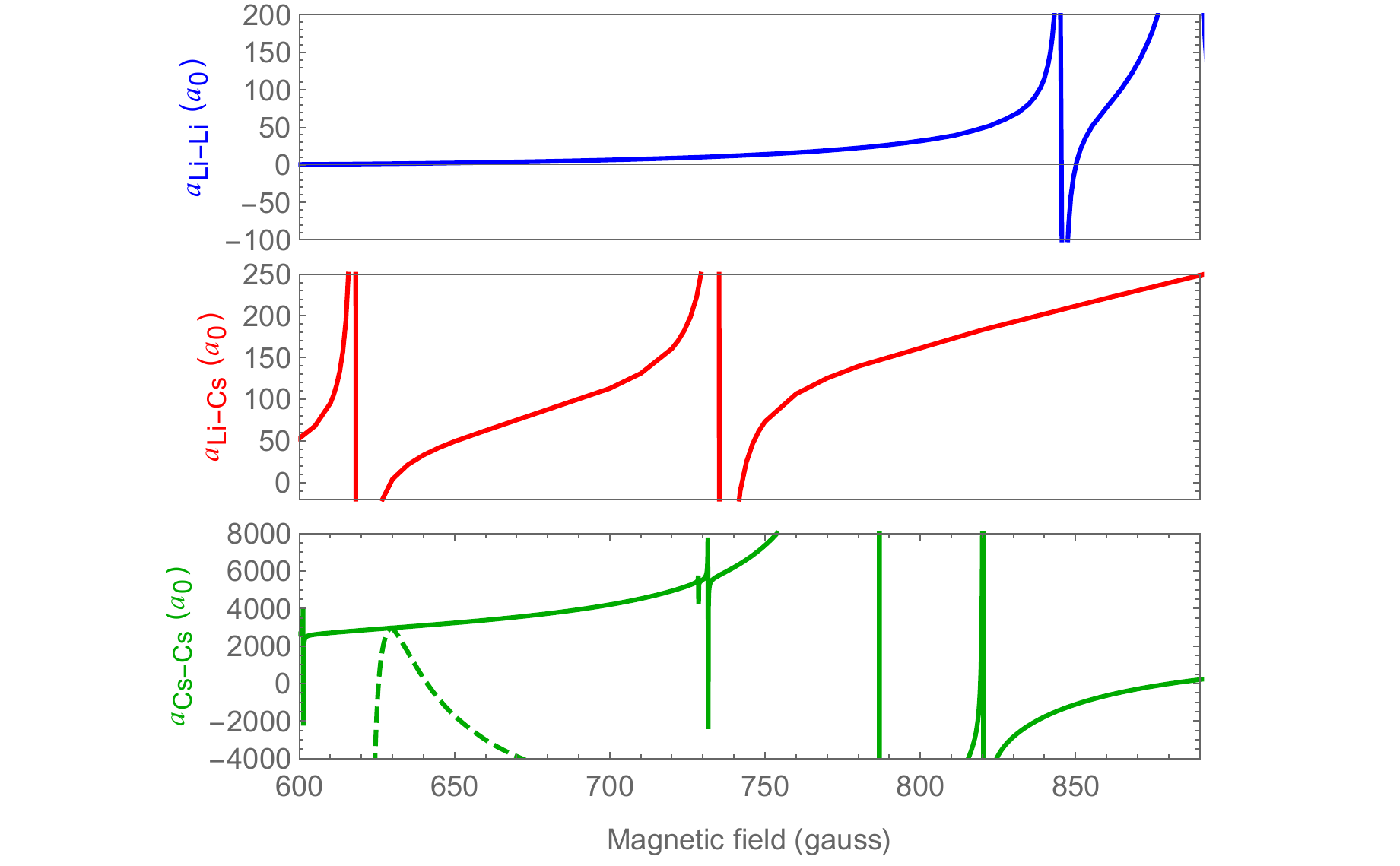}
\par\end{centering}
\caption{\label{fig:Cases}Scattering lengths of a mixture of lithium-7 and
caesium-133 atoms as a function of magnetic field, with lithium-7
in the ground hyperfine state $a$ (left panel) and the first excited
hyperfine state $b$ (right panel). The Cs-Cs scattering length is
taken from Ref.~\cite{Berninger2013}, while the Li-Li scattering
length is taken from Ref.~\cite{Pollack2009a} (left) and Ref.~\cite{Shotan2014}
(right). The dashed curve represents the effective scattering length
of caesium impurities immersed the lithium-7 condensate, obtained
from the formula Eq.~(\ref{eq:EffectiveScatteringLength}). }

\end{figure*}

The scattering length and near-threshold bound states are plotted
in Fig.~\ref{fig:aa} for the $aa$ channel. One can see an extremely
broad resonance occurring near $B=3000$~G. This situation is attributed
to the anomalously large triplet scattering length, resulting in a
large negative background scattering length. It is likely that a virtual
state very close to the threshold of the open channel shifts and broadens
the resonance. A similar situation occurs in $^{6}$Li-$^{6}$Li scattering~\cite{Chin2010}.
It can be checked that artificially changing the triplet scattering
length to a more nominal value (on the order of $\bar{a}$) results
in a much narrower resonance at smaller magnetic field, very close
to the first resonance in Fig.~\ref{fig:aa} around 538~G.

The results for the $ab$ channel are shown in Fig.~\ref{fig:ab}
in the range of magnetic field where collisions are elastic. It turns
out that no resonance appears in this range. The scattering length
is negative and varies from -100~$a_{0}$ to zero. For larger magnetic
fields, in the $ba$ channel (Fig.~\ref{fig:ba}), a situation similar
to the $aa$ channel occurs. Two resonances are found around 618 and
735~G, in addition to an extremely broad resonance occurring near
$B=3000$~G, resulting again from the anomalously large triplet scattering
length. The broad resonance creates again a wide range of magnetic
field where the scattering length varies linearly. 

Each resonance can be fitted by the usual formula,
\begin{equation}
a=a_{bg}-\frac{\Delta}{B-B_{0}}\label{eq:ResonanceFormula}
\end{equation}
where $a_{\text{bg}}$ is the background scattering length, $B_{0}$
is the resonance position, and $\Delta$ is the magnetic width of
the resonance. The resonances and their parameters are summarised
in Table.~\ref{tab:Positions7Li-133Cs}\@.

\section{Discussion}

In light of these results, we can look into whether the predicted
resonances could be used to investigate Efimov and polaron physics.
One aspect to consider is the interaction between atoms of the same
species. In particular, for lithium-7 to form a stable condensate,
the Li-Li scattering length must positive and not too large. A negative
scattering implies that the condensate of lithium-7 would be either
unstable, or limited to a very small number of atoms. 

In the lithium-7 $a$ state, the Li-Li scattering length is positive
only at magnetic fields $B<140$~G and in the window 540~G~$<B<$~740~G~\cite{Pollack2009a}.
In the first region, the Li-Cs scattering length is nearly constant
and negative, around $-150\,a_{0}$. Although this could not be used
to study resonant physics, it could be used to study attractive Bose
polarons (i.e. impurities interacting attractively with a condensate).
In the second region, shown in the left panel of Fig.~\ref{fig:Cases},
the Li-Cs scattering length grows nearly linearly from nearly 0 to
about $200\,a_{0}$. Again, this region cannot be used to study resonant
physics, but repulsive Bose polarons may be produced (i.e. impurities
interacting repulsively with a condensate). Although the direct
interaction between caesium atoms is relatively strong (with a scattering
length around 3000~$a_{0}$), their scattering length is affected
by the interaction induced by lithium-7 atoms. The resulting effective
scattering length can be estimated from the simple formula~\cite{Pethick2002},
\begin{equation}
\tilde{a}_{\text{Cs-Cs}}=a_{\text{Cs-Cs}}-\frac{Mm}{4\mu^{2}}\frac{a_{\text{Li-Cs}}^{2}}{a_{\text{Li-Li}}}\label{eq:EffectiveScatteringLength}
\end{equation}
where $M$ is the mass of caesium-133, $m$ is the mass of lithium-7,
and $\mu=(1/m+1/M)^{-1}\approx m$ is their reduced mass. This effective
scattering length is shown as a dashed curve in Fig.~\ref{fig:Cases},
where it can be seen to vanish around 650~G.

In the lithium-7 $b$ state, the Li-Li scattering length is positive
in larger regions of magnetic field, namely $B<400$~G and 600~G~$<B<$~800~G.
The first region cannot be used, since the lithium-caesium $ba$ channel
is inelastic in that region. The second region is more promising for
resonant physics, as it includes the two $ba$ resonances at 618~G
and 735~G. However, both resonances are relatively narrow and require
a precise stabilisation of the magnetic field. Moreover, the Cs-Cs
scattering length around these resonances is relatively large (around
4000~$a_{0}$) which may cause unwanted losses if the density of
caesium atoms is too large. We note that the effective Cs-Cs scattering
length may again be reduced near 650~G. Alternatively, one could
work around 870~G where the direct Cs-Cs scattering length vanishes,
however in that region the Li-Cs scattering length may not be varied
by much and the Li-Li scattering length in turn becomes near-resonant.

\section{Conclusion}

It was found that the bosonic lithium-caesium mixtures exhibit a few
interspecies resonances accessible to experiments. While these resonances
may not be ideal for studying Efimov physics, the lithium-caesium
mixtures nevertheless present some interesting opportunities for studying
Bose polaron physics.\vspace{0.5cm}

\noindent \rule[0.5ex]{1\columnwidth}{1pt}

The author thanks Bing Zhu for suggesting these calculations and his
helpful comments. The authors also thanks Paul Julienne for providing
the caesium scattering length data of Ref.~\cite{Berninger2013}.
This work was supported by the RIKEN Incentive Research Project and
JSPS Grants-in-Aid for Scientific Research on Innovative Areas (No.
JP18H05407).

\bibliographystyle{IEEEtran2}
\bibliography{paper33}

\end{document}